\documentclass[iop,apj,tighten]{emulateapj}
\usepackage{apjfonts} 
\usepackage{amsmath,amstext}
\usepackage{graphicx}
\usepackage{rotating}
\usepackage{multirow}
\usepackage{verbatim}
\usepackage{threeparttable}
\usepackage[breaklinks,colorlinks,citecolor=blue,linkcolor=magenta]{hyperref} 
\usepackage[all]{hypcap} 

\begin{document}

\title{Characteristics of ephemeral coronal holes}
\author{A. R. Inglis\altaffilmark{1,2}, R. E. O'Connor\altaffilmark{3,4}, W. D. Pesnell\altaffilmark{1}, M. S. Kirk\altaffilmark{1,2} and N . Karna\altaffilmark{5}}
\affil{1. Solar Physics Laboratory, Heliophysics Science Division, NASA Goddard Space Flight Center, Greenbelt, MD, 20771, USA}
\affil{2. Physics Department, The Catholic University of America, Washington, DC, 20064, USA}
\affil{3. Ball Aerospace, Boulder, CO, 80301, USA}
\affil{4. Smith College, Northampton, MA, 01063, USA} 
\affil{5. Harvard-Smithsonian Center for Astrophysics, Cambridge, MA, 02138, USA}

\begin{abstract}
 Small-scale ephemeral coronal holes may be a recurring feature on the solar disk, but have received comparatively little attention. These events are characterized by compact structure and short total lifetimes, substantially less than a solar disk crossing. We present a systematic search for these events, using Atmospheric Imaging Assembly EUV image data from the Solar Dynamics Observatory, covering the time period 2010 -- 2015. Following strict criteria, this search yielded four clear examples of the ephemeral coronal hole phenomenon. The properties of each event are characterized, including their total lifetime, growth and decay rates, and areas. The magnetic properties of these events are also determined using Helioseismic and Magnetic Imager data. Based on these four events, ephemeral coronal holes experience rapid initial growth of up to $\sim$3000 Mm$^2$ hr$^{-1}$, while the decay phases are typically more gradual. Like conventional coronal holes, the mean magnetic field in each ephemeral coronal hole displays a consistent polarity, with mean magnetic flux densities generally < 10~G. No evidence of a corresponding signature is seen in solar wind data at 1~AU. Further study is needed to determine whether ephemeral coronal holes are under-reported events or a truly rare phenomenon.

\end{abstract}

\keywords{Sun: corona --- Sun: magnetic fields --- Sun: UV radiation}
\maketitle

\section{Introduction}
\label{introduction}

Coronal holes are low density regions of the solar corona, typically observed as dark features at EUV wavelengths, and are associated with areas where the Sun's magnetic field opens into space and the interplanetary medium \citep{2009LRSP....6....3C}. These areas of open field allow the solar wind to propagate at high speed into space and towards the Earth \citep[e.g.][]{1972SoPh...26..354A, 1999Sci...283..810H, 2007SoPh..241..371T}. This open-field nature also means that they are usually dominated by a single magnetic polarity. Coronal holes typically occupy between 5 - 20\% of the solar disk depending on the level of solar activity \citep{2009SSRv..144..383W}, generally covering the least area during the maxima of the solar activity cycle. The latitudes at which coronal holes are found is also dependent on the solar cycle; at solar minimum coronal holes are confined to higher latitudes and typically cover the polar areas of the solar disk, while at solar maximum they can appear at more equatorial latitudes \citep{2011SSRv..158..267B}. Coronal holes can also be distinguished by their spectral abundance signatures. Unlike other coronal regions such as active regions and quiet Sun areas, the abundances of low first ionization potential (FIP) elements in coronal holes are very similar in the upper chromosphere and coronal layers compared with the corresponding photospheric abundances \citep{2003SSRv..107..665F}. 

The term coronal hole encompasses a full range of observationally similar phenomena, from large, long-lived and stable polar coronal holes \citep[e.g.][]{2009SoPh..257...99K, 2014SoPh..289.4047H, 2014SoPh..289.3381K}, to shorter-lived, more rapidly evolving equatorial coronal holes \citep[e.g.][]{2018AJ....155..153K}, and may also refer to low density off-limb regions \citep{2009LRSP....6....3C}. Shorter-lived equatorial coronal holes are often considered as the cause of moderate geomagnetic storms \citep{2011A&A...533A..49V}. A physically different phenomenon is a coronal dimming \citep{2008ApJ...674..576R, 2017ApJ...839...50K}, despite being also known as a `transient coronal hole'; these events are associated with the evacuation of plasma from a region in the aftermath of eruptive events, i.e. flares and coronal mass ejections (CMEs).

In this work we focus on compact, short-lived `ephemeral' coronal holes (EphCH), an observational subclass of the coronal hole phenomenon. These structures are distinct from longer-lived equatorial and polar coronal hole regions, as well as from coronal dimmings. Ephemeral coronal holes are primarily characterized by a lifetime substantially less than a single solar disk crossing, typically lasting only a few days, but are not associated with eruptive events. Within that timeframe, they experience rapid growth and dissolution phases, while remaining compact compared to conventional equatorial or polar coronal holes. These ephemeral coronal holes have been paid relatively little attention, and their rarity remains uncertain.

Here, we identify and characterize four examples of the ephemeral coronal hole phenomenon using high-resolution imaging data from the Atmospheric Imaging Assembly \citep[AIA;][]{2012SoPh..275...17L} and the Helioseismic and Magnetic Imager \citep[HMI;][]{2012SoPh..275..207S} on board the Solar Dynamics Observatory \citep[SDO;][]{2012SoPh..275....3P}. In Section \ref{methods} we describe the search for ephemeral coronal holes and the methodology for their characterization. In Section \ref{results} we describe in detail the properties of the 4 coronal holes studied in detail. The results and implications are summarized in Section \ref{discussion}.

\section{Methodology}
\label{methods}

\subsection{Finding coronal holes}
\label{finding_holes}
The first ephemeral coronal hole to be identified (EphCH) was the event of 2010 October 26 (see Section \ref{20101026}). Following this, a manual search of SDO/AIA image data between 2010 -- 2015 was conducted in order to find further similar events. This was achieved by sequential manual inspection of 211~\AA\ and 193~\AA\ images using the Helioviewer\footnote{\url{https://helioviewer.org/}} tool in search of candidate isolated dark regions.  Several criteria were established in order to identify potential EphCH events. First, a candidate event must be a clearly darkened, discrete region of the solar corona. Second, the darkened structure should be relatively isolated and free from interference from external phenomena. The third criterion was that the full lifetime of darkened structure must be less than the time taken to traverse the visible solar disk, i.e. the entire lifetime of the coronal hole must be less than $\sim$ 14 days. Finally, the identified feature should not be disrupted by an eruptive event during its lifetime, in order to distinguish them from coronal dimmings.

Following these criteria, 151 potential EphCH were selected for closer examination. Of these, we identify and examine in detail the four clearest examples of the EphCH phenomenon, including the originally-discovered 2010 October 26 event. These events are summarized in Table \ref{ephch_properties}. For the remainder of the sample, upon closer examination 112 candidate events were clearly unsuitable for analysis based on at least one of the following criteria: insufficient darkening (10 candidate events), lifetime extending off disk (63 events), or not isolated (39 events). A further 35 events were considered to be unclear for similar reasons and not studied further: lifetime extending off disk (19 events), insufficient darkening (8 events), unusual morphology (2 events), or a nearby disruptive eruption (6 events). These criteria were intentionally strict for the purpose of identifying only the clearest events for study.

\subsection{Data selection}

For each of the four ephemeral coronal hole examples, we obtain partial field-of-view level 1.5 SDO/AIA 211~\AA\ images at a 1-hour cadence using the SDO cutout service\footnote{\url{https://www.lmsal.com/get_aia_data/}}, covering the lifetime of each event. This wavelength was chosen because coronal holes are often most easily observed in the 211~\AA\ and 193~\AA\ channels. For the same time window, we acquire level 1.5 SDO/HMI images of the line-of-sight magnetic field, at a nominal 1-hour cadence.

During the 2010 October 26 event (EphCH 1), SDO is undergoing roll and calibration maneuvers, meaning that at some time intervals data is either not available, or the image is distorted. For example, the image obtained at 13:00 UT on 2010 October 26 is severely distorted, and is discarded from subsequent analysis. For HMI data, several image frames were unavailable due to maneuvers; in particular, between the period 12:00 UT and 18:00 UT on 2010 October 26, only one HMI magnetic field image was calculated, at 16:00 UT. For the other studied coronal holes, HMI frames are also occasionally unavailable. For all of the studied coronal holes, when HMI images are unavailable at times coincident with the AIA observations, the image frame closest to the AIA time is used to estimate the magnetic field within the coronal hole.

\subsection{Characterizing coronal holes}
\label{measure_size}

A primary characterization task for ephemeral coronal holes is to estimate their shape and extent as a function of time over their entire lifetime. We achieve this as follows:

\begin{enumerate}
\item For a given ephemeral coronal hole, we create a time-ordered stack of SDO/AIA 211~\AA\ images at 1-hour cadence covering the solar region of interest. These images are each associated with a time $t_i$, for $i = 1,2,3...n$, where $n$ is the number of images in the stack.

\item In order to reduce image noise, each image in the stack is smoothed in two dimensions, with a smoothing scale encompassing each pixel's immediate neighbours. This smoothing scale is small relative to the coronal structures of interest.

\item for each time $t_i$ in the image stack, the pixel corresponding to the minimum intensity in the region of interest is identified. This is defined as the origin point $O(x,y)$ for that time.

\item From the origin point $O(x,y)$, a region growth algorithm is used to expand in all directions until a threshold pixel intensity value is reached. This is achieved using the REGION\_GROW function in IDL. This function finds all pixels that are connected (via pixel edges or diagonals) to the origin point that do not violate the threshold conditions, i.e. every pixel in the final region has a path to the origin point via other pixels that are within the intensity threshold. This approach is convenient for defining regions of irregular shape. In this work, a fixed threshold intensity value is chosen, with a correction for the loss of sensitivity of the AIA channels over time \citep{2014SoPh..289.2377B}. This sensitivity correction was calculated using the \verb|aia_get_response.pro| function within the SolarSoftWare package, and finding the total effective area of each AIA channel as a function of time. The threshold value is $F_{thresh}$ = 15~DN s$^{-1}$ for the initial 2010 October 26 event. Using the function described above, we can find the appropriate correction coefficients relative to EphCH 1 for EphCH 2, 3 and 4. These sensitivity correction factors for the 211~\AA\ channel are found to be 0.90, 0.83 and 0.77 respectively. 

\item Finally, a morphological closing operation \citep[see, e.g.][]{2002dip..book.....G} is applied to the grown region, using a 5x5 pixel scale. This `closing' effectively fills small gaps in the defined region. The area of the image encompassed by this processed grown region is considered to be the coronal hole area for the given $t_i$. 
\end{enumerate}

The advantage of this approach is that it requires the coronal hole region to be contiguous, but does not put other constraints on the morphology of the coronal hole. Using a fixed threshold value $F_{thresh}$ with time-dependent sensitivity correction to define the extent of each coronal hole allows us to easily compare the properties of the ephemeral coronal holes studied in this paper, despite them occurring in different calendar years with changing instrument performance. The threshold value is also consistent with those derived from slightly different methodologies. For example, during a study of a long-lived low latitude coronal hole, \citet{2018ApJ...861..151H} adopted a threshold of 35\% of the median intensity values in the image. For the events studied here, the median intensity values for 211~\AA\ on-disk pixels across the 4 studied time intervals was $\sim$ 40 DN s$^{-1}$, hence adopting this approach would yield a threshold of $\sim$ 14 DN s$^{-1}$. For studies of larger samples of events, an automated detection approach may be more advantageous \citep[e.g.][]{2009SoPh..257...99K, 2009SoPh..256...87K, 2014A&A...561A..29V, 2018SoPh..293...71H}.

In order to accurately estimate the area encompassed by the ephemeral coronal hole, we correct for projection effects due to the curvature of the solar disk. As a result of this projection, pixels near disk center represent a smaller physical area than those nearer the solar limb. This correction is given for pixel $j$ by \citep[e.g.][]{2018ApJ...861..151H}:

\begin{equation}
A_{j, corr} = \frac{A_j}{\cos \alpha_j}
\label{area_eqn}
\end{equation}

where $A_j$ is the uncorrected area of pixel $j$, and $\alpha$ is the angular distance from pixel $j$ to the disk center. The total de-projected area of the coronal hole can then be found at any $t_i$ by summing over all $A_j$ within the coronal hole boundary at that time.

\begin{figure}
    \centering
    \includegraphics[width=8cm]{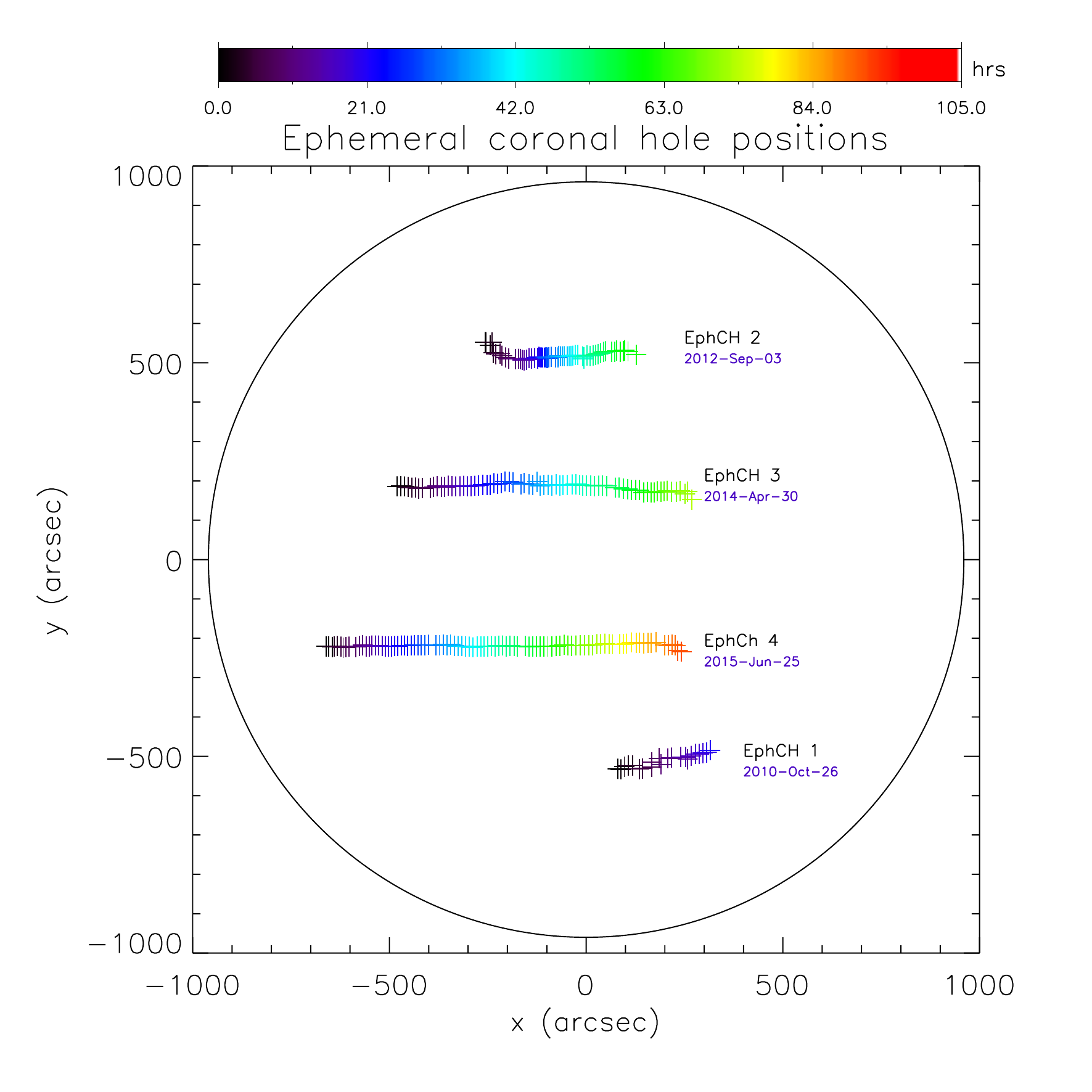}
    \caption{The approximate locations of each ephemeral coronal hole as a function of time, calculated using the centroid position of the coronal hole image mask. This centroid is distinct from the location of minimum flux identified in Section \ref{measure_size}. The color of the location markers represents the time elapsed since the ephemeral coronal hole was first visible. The start dates for each coronal hole are shown as labels next to each location track.}
    \label{tracks}
\end{figure}

The methodology described above allows us to obtain the position and shape of each coronal hole throughout its lifetime. The approximate locations of each ephemeral coronal hole on the solar disk are shown as a function of time in Figure \ref{tracks}, where the color of the EphCH location marker represents the time elapsed since it was first visible, $t_0$. Due to the selection criteria imposed in Section \ref{finding_holes}, these events exist on the solar disk and away from the solar limb. This is relevant as although the correction applied by Equation \ref{area_eqn} is appropriate for most locations on the solar disk, it can be problematic for pixels near the limb, where the foreshortening effect is so large that the estimated area can explode to very large, unrealistic values. 

Using the estimated area masks of the ephemeral coronal holes as a function of time, we find the properties of the magnetic field within the hole using co-aliged HMI images (see Figure \ref{examples_all}). We also estimate the growth and decay rates for each event. This is done via linear fits to the total corrected area during manually defined growth and decay phases for each event (see e.g. Figure \ref{area_and_flux_20101026}). This approach does not attempt to capture the full complexity of the coronal hole evolution, instead providing a first order estimate of growth and decay rates.

\section{Results}
\label{results}

\subsection{Ephemeral coronal hole 1: 2010-10-26}
\label{20101026}

\begin{figure*}[ht]
\begin{center}
\includegraphics[width=16cm]{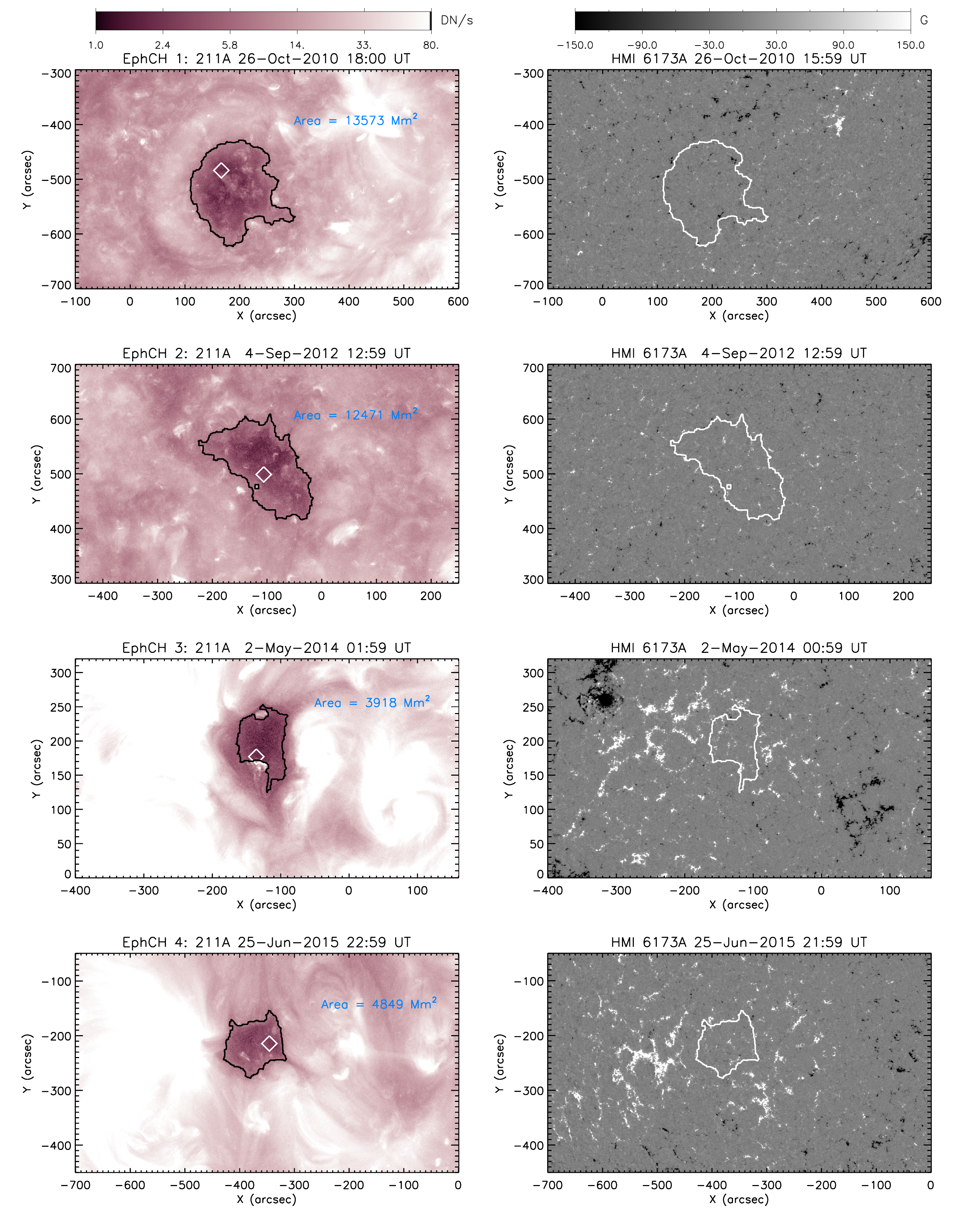}
\caption{SDO/AIA 211\AA\ and SDO/HMI images of each of the ephemeral coronal holes identified. Top row: The first ephemeral coronal hole (EphCH 1) from 2010 October 26 observed at 18:00 UT. The estimated coronal hole boundary is shown by the irregular black contour, while the origin point $O(x,y)$ used for determining this area is denoted by the white diamond (see Section \ref{measure_size}). The total de-projected area of the coronal hole is shown in blue text in the upper right. Second row: The second ephemeral coronal hole (EphCH 2) observed on 2012 September 4 at 13:00 UT. Third row: The third coronal hole (EphCH 3) observed at 2014 May 2 at 02:00 UT. Fourth row: The fourth coronal hole (EphCH 4) observed at 2015 June 25 at 23:00 UT. In these images, color scales have been clipped for clarity. }
\label{examples_all}
\end{center}
\end{figure*}

The first example of an ephemeral coronal hole was observed beginning on 2010 October 26. It first became visible around 05:00 UT on 2010 October 26, rapidly growing to reach its maximum area at 12:00 UT, before beginning to decay in size, becoming unobservable at $\sim$ 06:00 UT on 2010 October 27. Hence, the lifetime of this ephemeral coronal hole was only $\sim$ 24 hours. During this time, SDO was performing calibration maneuvers, resulting in some image frames for both AIA and HMI being unavailable. 

Figure \ref{examples_all}a shows a SDO/AIA 211\AA\ snapshot of the coronal hole extent at 2010-10-26 18:00 UT, as determined using the methodology described in Section \ref{measure_size}. In each image, the origin point $O(x,y)$ is denoted by the white diamond, while the black contour shows the coronal hole mask. For this event, a small active region is present to the solar northwest of the coronal hole. There is also evidence of a filament structure bounding the coronal hole on the eastern side.

\begin{figure}
\begin{center}
\includegraphics[width=8.5cm]{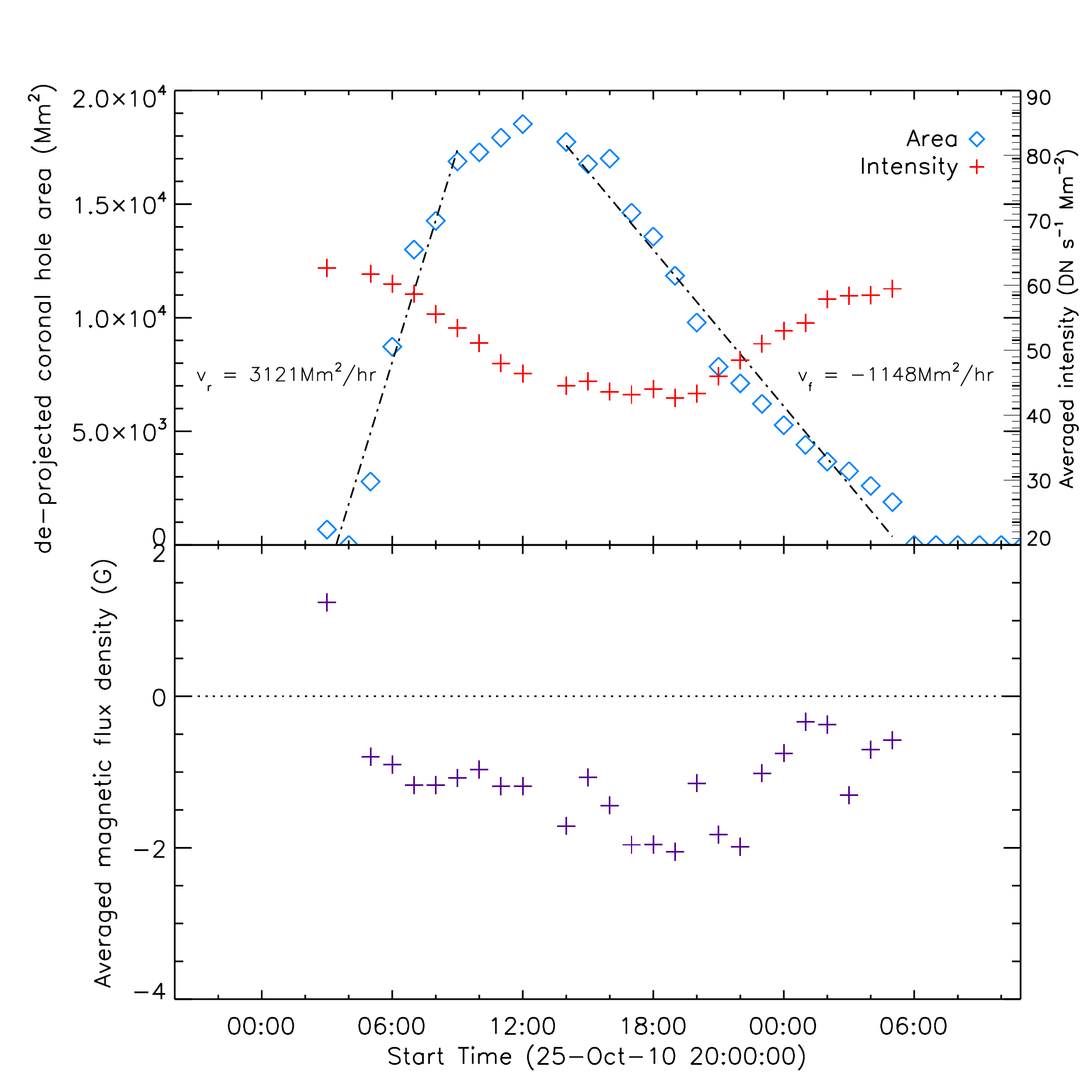}
\caption{Evolution of the properties of EphCH 1, from 2010 October 26. Top: The total de-projected area of the coronal hole as a function of time (blue diamonds), and the area-averaged intensity within the coronal hole boundary over time(red crosses). Bottom: The area-averaged magnetic flux density within the coronal hole boundary as a function of time, as measured by SDO/HMI.}
\label{area_and_flux_20101026}
\end{center}
\end{figure}

The total de-projected coronal hole area is shown as a function of time in Figure \ref{area_and_flux_20101026}, along with the AIA intensity per unit de-projected area within the coronal hole. From this we can see that the coronal hole experiences rapid growth from its onset at 06:00, reaching a maximum area of $\sim$ 1.9$\times$10$^{4}$ Mm$^2$ at 12:00 UT, before experiencing a more gradual decay phase, ending with dissolution of the coronal hole at $\sim$ 06:00 UT on 2010 October 27. The expansion rate is estimated to be $v_r$ $\sim$ 3100 Mm$^2$ hr$^{-1}$, while the decay rate following the peak is three times slower, $v_f$ $\sim$ -1150 Mm$^2$ hr$^{-1}$. During the expansion phase, the area-averaged \textbf{intensity} decreases by around $\sim$ 25\%, from 60 DN s$^{-1}$ Mm$^{-2}$ to a minimum of $\sim$ 45 DN s$^{-1}$ Mm$^{-2}$. Hence, during this time, the area of reduced EUV intensity not only expands, but the individual pixels within the coronal hole also continue to darken. This minimum flux level lasts well into the decay phase of the coronal hole; the area-averaged flux does not begin to rise again until $\sim$ 20:00 UT, 8 hours after the beginning of the coronal hole decay phase. After 20:00 UT, the coronal hole flux gradually rises again, reaching its starting level as the hole dissipates.

Figure \ref{area_and_flux_20101026} also shows the average magnetic flux density in the coronal hole, as measured by HMI. These magnetic field measurements are corrected for each pixel's angle to disk center, assuming a radial field. This shows that the coronal hole is situated in an area of weak, slightly negative magnetic field. There is also some indication that the average flux density increases in tandem with the darkening effect observed in the EUV intensity. However, this should be treated with caution; roll manoeuvres taking place during the observations introduce an additional source of uncertainty for these magnetic field measurements.

\subsection{Ephemeral coronal hole 2: 2012-09-03}

The second ephemeral coronal hole to be observed began on 2012 September 3. Compared to EphCH 1, this event was substantially longer lived, with a total lifetime from first appearance to dissolution of $\sim$ 3 days. Figure \ref{examples_all}b shows an example of the EphCH 2 extent, at 13:00 UT on 2012 September 4.

\begin{figure}
\begin{center}
\includegraphics[width=8.5cm]{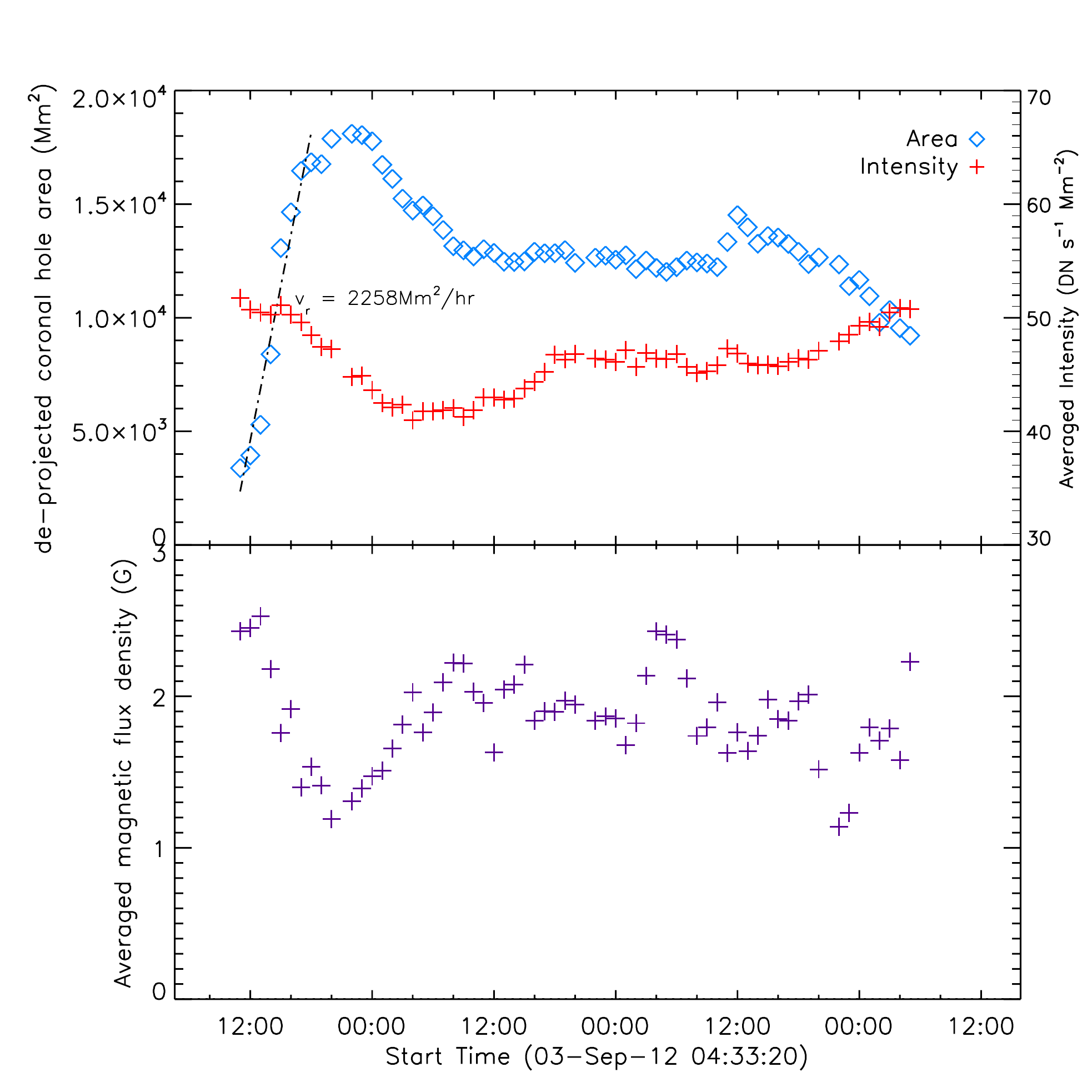}
\caption{Evolution of the properties of EphCH 2, from 2012 September 3-6. Top: The total de-projected area of the coronal hole as a function of time (blue diamonds), and the area-averaged intensity within the coronal hole boundary over time(red crosses). Bottom: The area-averaged magnetic flux density within the coronal hole boundary as a function of time, as measured by SDO/HMI.}
\label{area_and_flux_20120903}
\end{center}
\end{figure}

As with EphCH 1, this event appears and grows quickly to its maximum extent, reaching an estimated de-projected area of $\sim$ 1.8$\times$10$^{4}$ Mm$^2$ approximately 6 hours after its first appearance (see Figure \ref{area_and_flux_20120903}, with an estimated growth rate of $v_r$ $\sim$ 2300 Mm$^2$ hr$^{-1}$. Over the next $\sim$ 16 hours, the area decays to around 1.5$\times$10$^4$ Mm$^2$, at which point shrinking stops. Following this, the coronal hole remains stable in size until its abrupt dissolution at $\sim$ 08:00 UT on 2012 September 6. As such, it was not possible to estimate a decay rate for this event. Initially, the average intensity from within EphCH 2 displays similar behaviour to that of EphCH 1; there is an initial decrease during the growth phase, until the average intensity level reaches a minimum significantly later than the time of maximum extent. Following this minimum, the average intensity displays a slight but steady increase, up until the dissolution of the coronal hole.

Figure \ref{area_and_flux_20120903} also shows that the magnetic flux density within EphCH 2 is somewhat anti-correlated with the de-projected hole area, with the average absolute value dropping as the area expands. This is most notable during the two peaks in the area of EphCH 2 at 18:00 UT on 2012 September 4, and at 12:00 UT on 2012 September 5. Quantifying this with a Pearson correlation, we find a correlation coefficient $C = -0.64$. This anti-correlated behaviour appears to be unique among the ephemeral coronal holes studied here; for EphCH 1, 3 and 4, the unsigned magnetic flux density is either correlated with the de-projected coronal hole area or approximately constant and independent of the de-projected area.

\subsection{Ephemeral coronal hole 3: 2014-04-30}

\begin{figure}
\begin{center}
\includegraphics[width=8.5cm]{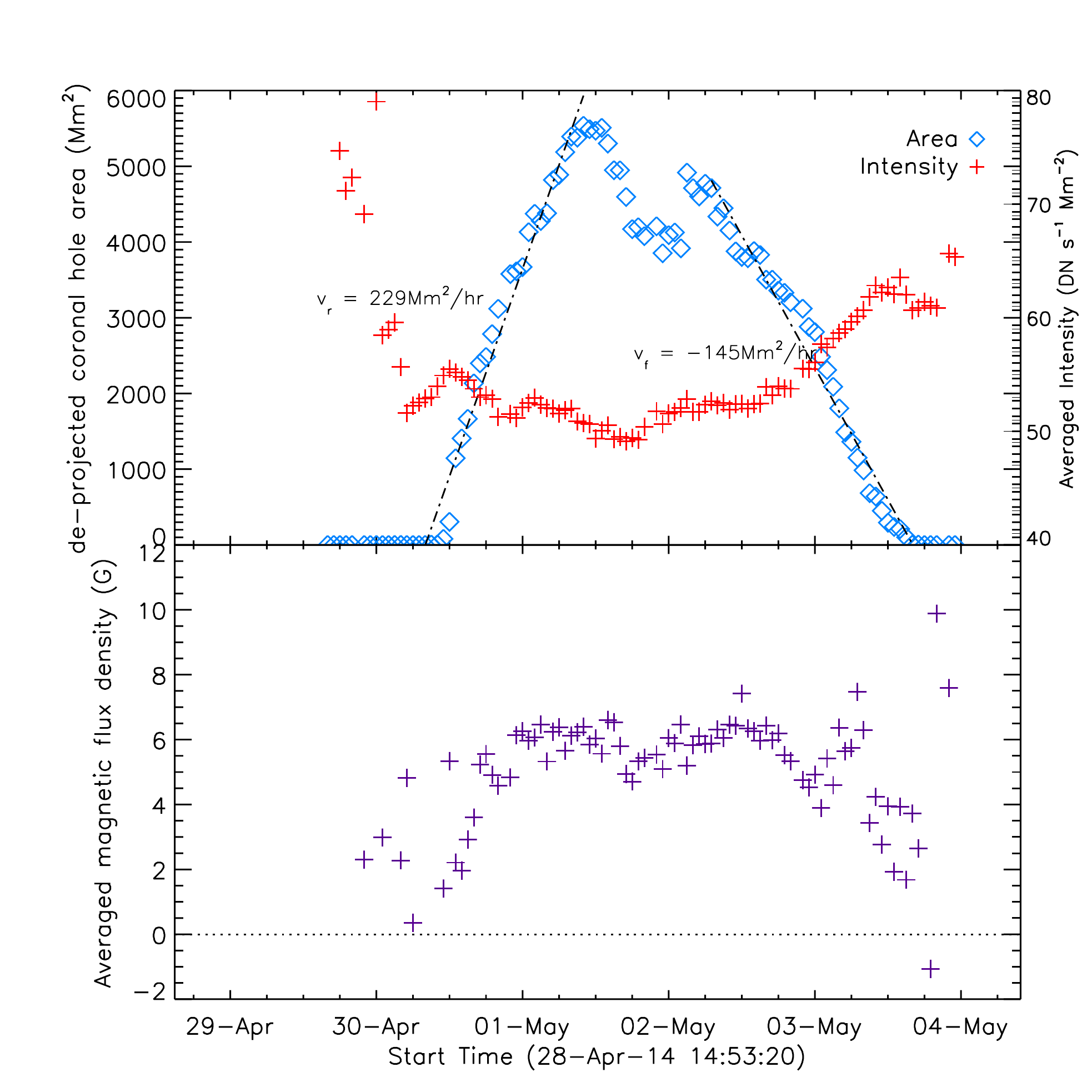}
\caption{Evolution of the properties of EphCH 3, from 2014 April 30 - 2014 May 4. Top: The total de-projected area of the coronal hole as a function of time (blue diamonds), and the area-averaged intensity within the coronal hole boundary over time(red crosses). Bottom: The area-averaged magnetic flux density within the coronal hole boundary as a function of time, as measured by SDO/HMI.}
\label{area_and_flux_20140429}
\end{center}
\end{figure}

The third ephemeral coronal hole of this study was visible between approximately 2014 April 30 06:00 UT - 2014 May 04 00:00 UT, a total lifetime of $\sim$ 4 days (see Figure \ref{examples_all}c). Unlike EphCH 1 and EphCH 2, this event experienced a more gradual growth phase, not reaching its maximum extent until 12:00 UT on 2014 May 01, 30 hours after initial emergence. This evolution corresponded to an average growth rate of $\sim$ 230 Mm$^2$ hr$^{-1}$. EphCH 3 then experienced a dip and then recovery in area over the next 18 hours, before beginning a gradual decay phase at $\sim$ 06:00 UT on 2014 May 02 until its disappearance. The decay was at a somewhat slower rate than the initial growth, with an estimated decay rate of $\sim$ $-150$ Mm$^2$ hr$^{-1}$. The full evolution is shown in Figure \ref{area_and_flux_20140429}. The maximum de-projected area of EphCH 3 is substantially smaller than EphCH 1 and EphCH 2, reaching $\sim$ 5500 Mm$^2$.


For EphCH 3, the average magnetic flux density is stronger than that of EphCH 1 or EphCH 2 (see Figure \ref{area_and_flux_20140429}). It increases by a factor of three during the initial growth phase of the coronal hole, to $\sim$ 6 G. Thereafter, the magnetic flux density remains stable up until the dissolution of the coronal hole three days later.

\subsection{Ephemeral coronal hole 4: 2015-06-24}

\begin{figure}
\begin{center}
\includegraphics[width=8.5cm]{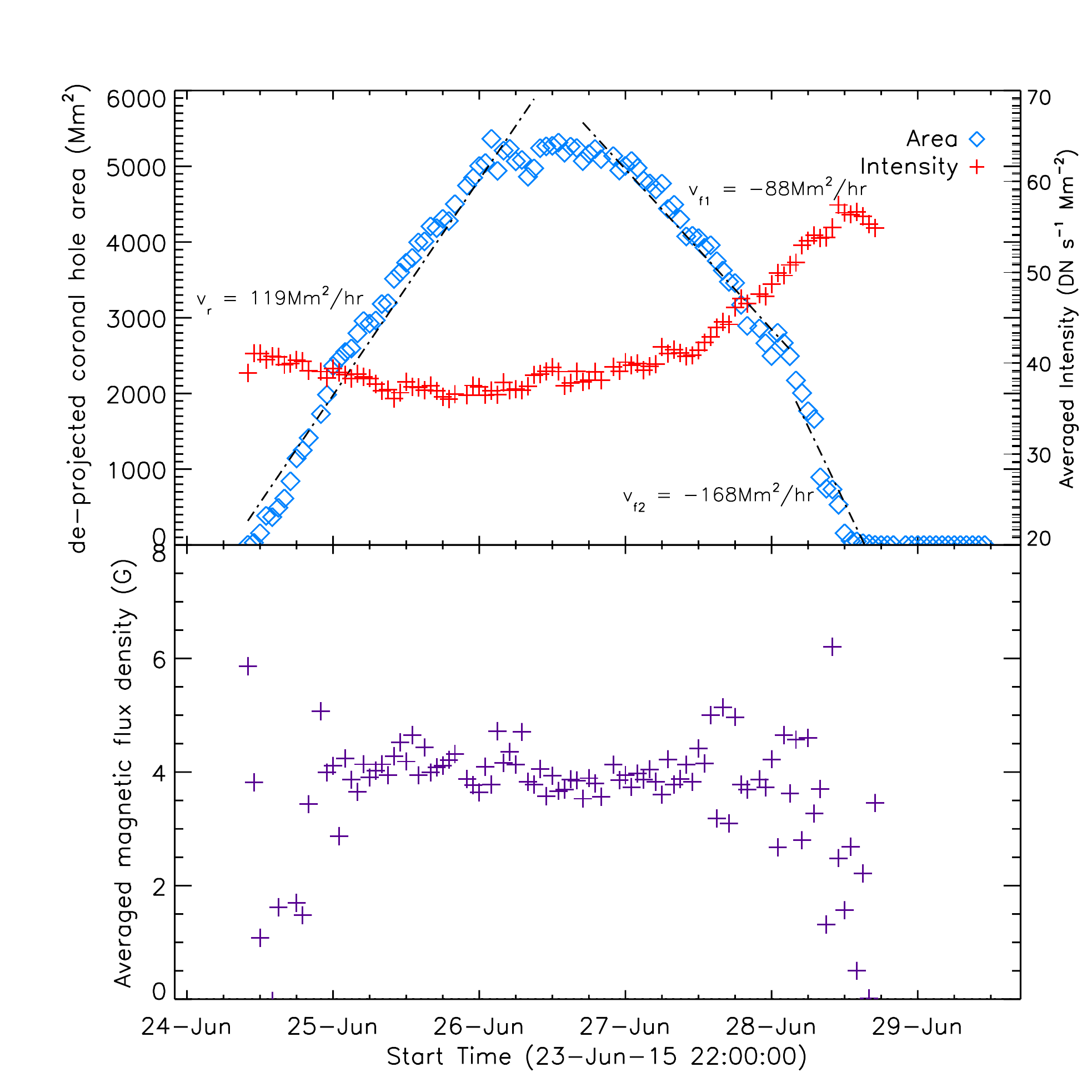}
\caption{Evolution of the properties of EphCH 5, from 2015 June 24 - 2015 June 28. Top: The total de-projected area of the coronal hole as a function of time (blue diamonds), and the area-averaged intensity within the coronal hole boundary over time(red crosses). Bottom: The area-averaged magnetic flux density within the coronal hole boundary as a function of time, as measured by SDO/HMI.}
\label{area_and_flux_20150625}
\end{center}
\end{figure}

The visibility of the final ephemeral coronal hole in this study spanned 2015 June 24 12:00 -- 2015 June 28 18:00, a total lifetime of more than 4 days. A Snapshot of this event is shown in Figure \ref{examples_all}d, indicating that EphCH 4 retained a relatively compact structure throughout. The evolution of EphCH 4 is charted by Figure \ref{area_and_flux_20150625}. The growth phase for this coronal hole lasted approximately 42 hours, plateauing at a maximum de-projected area of $\sim$ 5300 Mm$^2$ at $\sim$ 06:00 UT on 2015 June 26. The mean growth rate was $\sim$ 120 Mm$^2$ hr$^{-1}$, the slowest of the ephemeral coronal holes studied here. Area decay began $\sim$ 12 hours later, at 18:00 UT on 2015 June 26. Decay accelerated abruptly at $\sim$ 06:00 UT on 2015 June 28, resulting in the complete disappearance of EphCH 4 by 18:00 UT that day. Thus, two decay rates were measured for this event, an initial decay rate $v_{f1}$ $\sim$ $-90$ Mm$^2$ hr$^{-1}$, and $v_{f2}$ $\sim$ $-170$ Mm$^2$ hr$^{-1}$.

The average intensity within EphCH 4 displays slightly different behaviour from the previously discussed events. An initial decrease in average intensity occurs only during the first 24 hours of the coronal hole lifetime. Following this, the average intensity increases steadily within the hole, a process that begins while the area itself is still increasing. This trend continues up until the dissolution point.

The average magnetic flux density bounded by EphCH 4 remains approximately constant throughout the coronal hole lifetime, apart from the very early and late phases of the event, where the flux density has a large scatter due to uncertainties. The value of the magnetic flux density is strong relative to EphCH~1 and EphCH~2, though EphCH~3 remains the event bounding the strongest average magnetic field.

\subsection{Relationship between ephemeral coronal hole parameters}

\begin{figure*}
\begin{center}
\includegraphics[width=18cm]{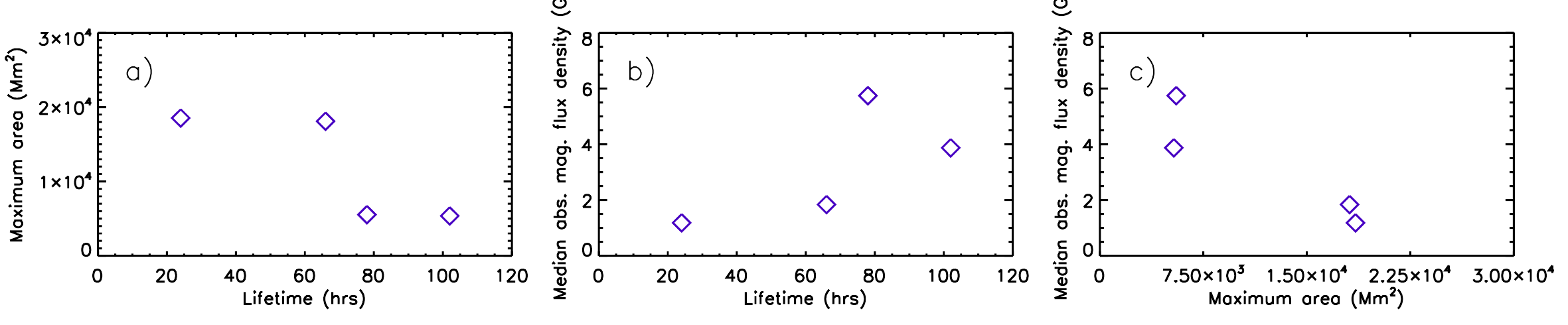}
\caption{Relationship between fundamental ephemeral coronal hole parameters. Panel a): Maximum de-projected coronal hole area reached versus the total coronal hole lifetime. Panel b): Median absolute magnetic flux density within the coronal hole versus coronal hole lifetime. Panel c): Median absolute magnetic flux density versus maximum de-projected coronal hole area.}
\label{correlations}
\end{center}
\end{figure*}

Given this selection of ephemeral coronal holes, we can investigate any links between their key fundamental parameters, including the de-projected area, lifetime, and magnetic flux density. Figure \ref{correlations} shows the relationship between these three properties.

With a sample size of only four events, it is not meaningful to perform a full correlation analysis of these parameters. However, some tentative observations may be drawn from Figure \ref{correlations}. First, Figure \ref{correlations}a shows that the two events with the longest lifetimes were more compact, with maximum areas $<$ 1000 Mm$^2$. These two compact coronal holes were also situated in regions with somewhat stronger median magnetic flux density (Figure \ref{correlations}b). Hence, by extension stronger flux density is also associated with smaller maximum area (Figure \ref{correlations}c). These shorter-lived coronal holes were also associated with much more rapid growth and decay rates compared with EphCH 3 and 4. Speculatively therefore, coronal holes sited in stronger field regions may be more stable, but more constrained in extent by the surrounding magnetic structure. However, a larger sample size is needed in order to ascertain whether these relationships are robust.

\subsection{Signatures in the solar wind}

Previous studies have shown that, like large coronal holes \citep[e.g.][]{2015SoPh..290.1355R}, small coronal holes can be associated with corresponding signatures in the solar wind \citep{2001JGR...10624915B}. This signature may be observed in a number of parameters including solar wind speed, temperature and density. Accordingly, we investigate the 1 AU solar wind data during the time intervals associated with each ephemeral coronal hole, starting from the time where the coronal hole crosses the central meridian. 

\begin{figure}
\begin{center}
\includegraphics[width=8.5cm]{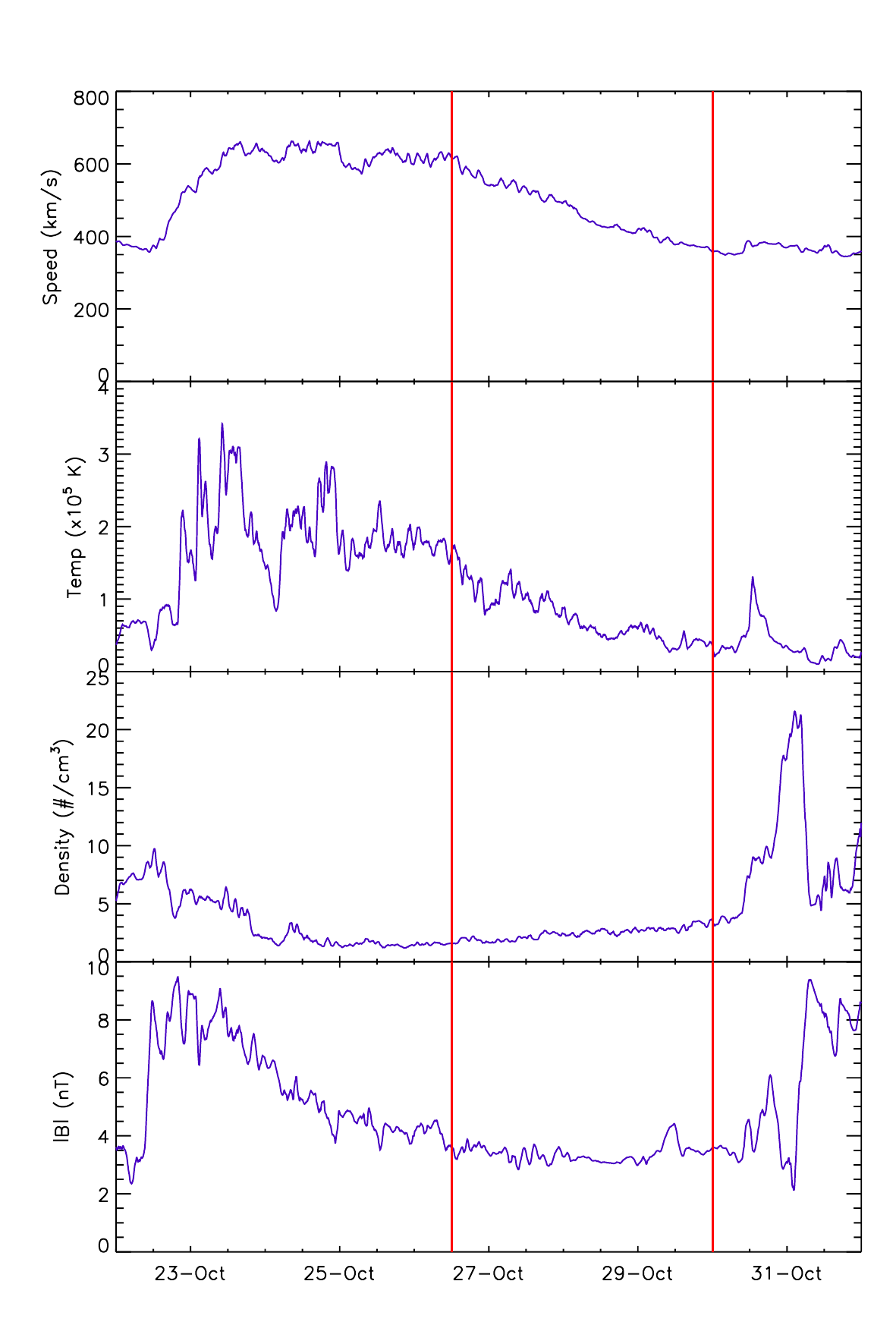}
\caption{Solar wind properties measured by WIND and obtained from the OMNI database during and following the lifetime of EphCH 1, including solar wind speed (first panel), temperature (second panel), density (third panel) and magnetic field strength (fourth panel). The first vertical line indicates the time at which EphCH 1 crossed the central meridian. The second vertical line corresponds to 3.5 days after the central meridian crossing.}
\label{solar_wind_20101026}
\end{center}
\end{figure}

Figure \ref{solar_wind_20101026} shows the 1 AU solar wind parameters associated with EphCH 1, beginning on 2010 October 21, as observed by the WIND spacecraft and incorporated into the OMNI database hosted by the Space Physics Data Facility (SPDF). The first vertical line shows the time at which EphCH 1 crosses the central meridian, while the second vertical line denotes 3.5 days since the central meridian crossing, the beginning of the time period at which a coronal hole signature in the solar wind would be expected to arrive at Earth \citep[e.g.][]{2015SoPh..290.1355R}. Such a signature would include an increase in the measured solar wind velocity \citep[e.g.][]{2015SoPh..290.1355R} and temperature, which may also be preceded by an initial rise in density and magnetic field $B$, although the density within the high-speed stream itself is generally low \citep[e.g.][]{1995JGR...10021717T, 2001JGR...10624915B, 2018JGRA..123.6457M}. As Figure \ref{solar_wind_20101026} shows, no clear signature is observable in the solar wind speed, density, temperature, or magnetic field strength at the expected arrival time. There is a barely-appreciable increase in $v_{sw}$ and only a brief spike in temperature, while the solar wind density does increase substantially, but around 4 days after central meridian crossing and after the observed changes in $v_{sw}$ and $T$. However, at the expected arrival time, the solar wind and temperature are already decaying from a previous moderate high-speed stream, which may mask any possible signature from the EphCH. Investigation of the remaining ephemeral coronal holes shows similar results. Although EphCH 2, 3 and 4 are all observed to cross the central meridian during their lifetimes, no discernible signature is observable in the 1 AU solar wind data near the expected Earth arrival times. This is not unexpected given their compact size and short lifetime relative to other coronal structures present at these times, as well as the varying connectivity between the Sun and the Earth.

\subsection{Comparison with large equatorial coronal holes}
\label{large_holes_comparison}

For context and for verification, it is useful to apply our methodology to SDO/AIA images of large, equatorial coronal holes. For this purpose we draw from the events studied by \citet{2018AJ....155..153K}, who focused on long-lived equatorial coronal holes that persisted for at least one solar rotation. This study characterized 190 CH events; here we briefly consider three representative examples at single points in their lifetimes, 1) 2011 August 24 12:00 UT, 2) 2011 September 9 00:00 UT and 3) 2011 April 10 00:00 UT respectively.

\begin{figure}
    \centering
    \includegraphics[width=9cm]{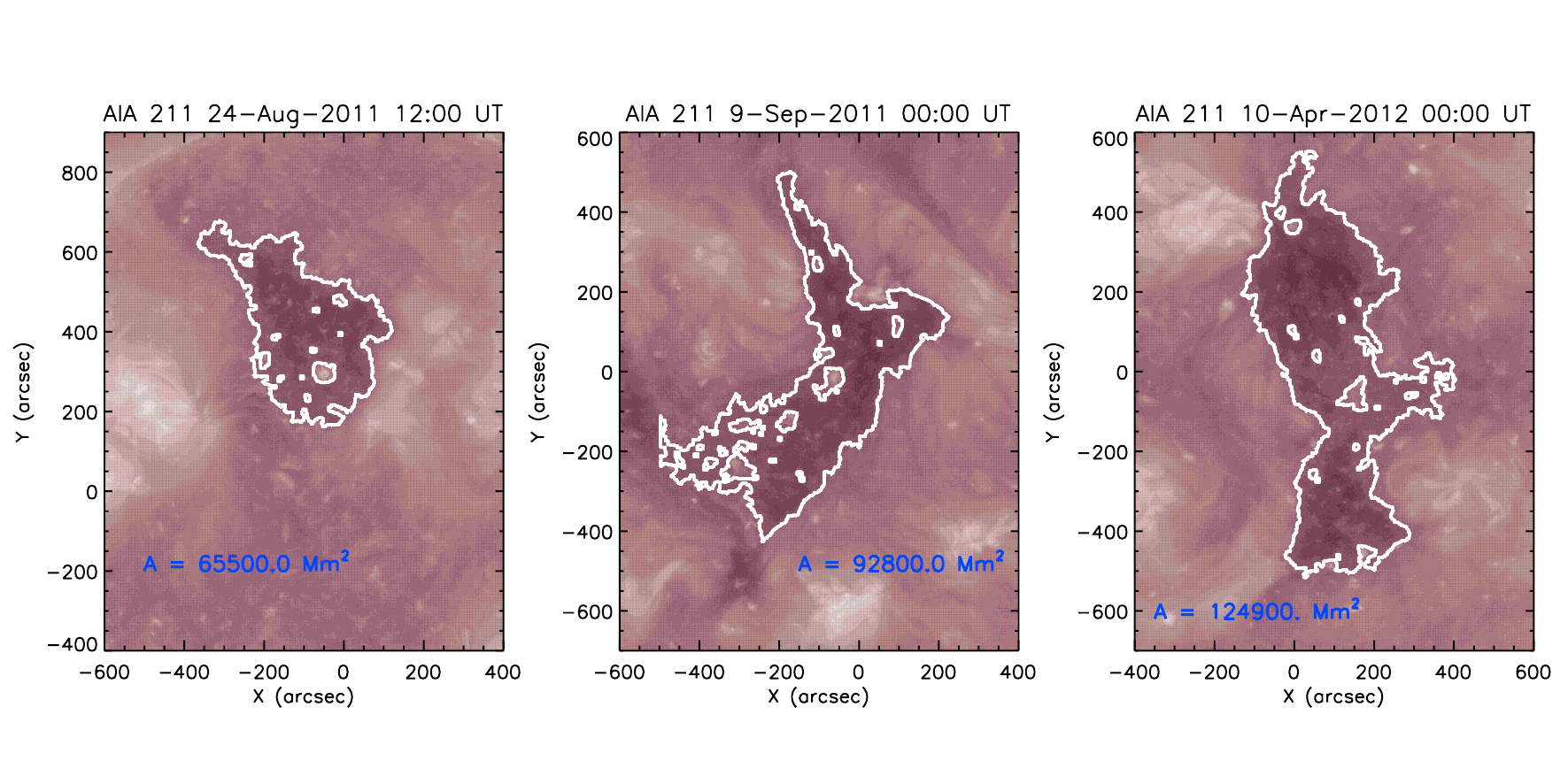}
    \caption{The coronal hole characterization methodology described in Section \ref{measure_size} applied to three long-lived equatorial coronal holes studied by \citet{2018AJ....155..153K}. Left: 2011 August 24 at 00:00 UT (SPoCA ID: 2380). Center: 2011 September 09 at 00:00 UT (SPoCA ID: 2453). Right: 2012 April 10 at 00:00 UT (SPoCA ID 4362). }
    \label{large_holes}
\end{figure}

As their lifetimes are longer than a solar rotation, we cannot guarantee observing these equatorial coronal holes at their maximum extent in the AIA 211 \AA\ channel as this may occur on the far side of the Sun. Similarly, we can not practically measure their growth and decay rates with our chosen methodology. Instead, we investigate the areas and magnetic flux densities of these events as they cross disk center, minimizing the area correction factors for these large structures and yielding a representative area. As with the ephemeral coronal holes, we follow the methodology described in Section \ref{measure_size}. Based on the occurrence times of these events, we apply the sensitivity correction factors of 0.95, 0.95 and 0.88 respectively to the threshold intensity value of 15 DN/s. The morphology of these equatorial coronal holes is shown in Figure \ref{large_holes}.

The areas of these equatorial coronal holes are much larger than the EphCH events that are the focus of this paper, with snapshot areas of $A_1$ $\sim$ $6.5\times10^4$ Mm$^2$, $A_2$ $\sim$ $9.3\times10^4$ Mm$^2$ and $A_3$ $\sim$ $1.25\times10^5$ Mm$^2$. The largest of these is approximately an order of magnitude larger than the ephemeral coronal holes at their maximum extent (see Table \ref{ephch_properties}). However, the underlying magnetic flux densities are similar, with average flux densities of $\sim$ -2~G, $\sim$ -2.75~G and $\sim$ -1.95~G respectively. 

We can also see from Figure \ref{large_holes} that our approach defines a realistic, credible boundary for these examples. This provides a useful verification of the methodology, showing that it is appropriate for estimating area and morphology not only for short-lived, compact EphCH structures, but also for large, long-lived equatorial coronal holes when they are not near the solar limb.

\section{Discussion and Conclusions}
\label{discussion}

\begin{table*}[]
\centering
\caption{Summary of ephemeral coronal hole (EphCH) properties}
\label{ephch_properties}
\begin{tabular}{c|c|c|c|c|c|c}
\hline
\hline
\# & Dates & Lifetime & Peak De-projected Area & Growth Rate & Decay Rate & Mean latitude \\
   &   & (hrs)          & (Mm$^2$) & (Mm$^2$/hr) & (Mm$^2$/hr) & (degrees) \\
\hline
1 & 2010-10-26 - 2010-10-27 & 24  & 1.9$\times$10$^4$ & 3.1$\times$10$^3$ & -1.1$\times$10$^3$ & $-31.3$ \\
2 & 2012-09-03 - 2012-09-06 & 66 & 1.8$\times$10$^4$ & 2.2$\times$10$^3$ & - & $33.4$ \\
3 & 2014-04-30 - 2014-05-04 & 78 & 5.5$\times$10$^3$ & 0.23$\times$10$^3$ & -0.15$\times$10$^3$ & $12.0$ \\
4 & 2015-06-24 - 2015-06-28 & 102 & 5.3$\times$10$^3$ & 0.12$\times$10$^3$ & -0.09$\times$10$^3$, -0.17$\times$10$^3$ & $-13.2$ \\

\end{tabular}
\end{table*}

We have identified and characterized the properties of four ephemeral coronal holes observed by SDO/AIA between 2010 -- 2015. These structures are a sub-class of the coronal hole phenomenon,  characterized by their lifetimes, compact size, isolation from other coronal features, and relatively rapid growth and decay. Coronal hole studies have traditionally focused on large equatorial and polar holes, rather than these small structures.

The peak areas for the 4 events studied were in the range 0.5$\times$10$^4$ -- 2.0$\times$10$^4$ Mm$^2$, showing that even at their maximum extent they remained compact structures. They also exhibited a range of growth and decay rates. Ephemeral coronal hole 1, observed on 2010 October 25 -- 27, was the shortest-lived of the events studied, and also experienced the most rapid growth and decay during its 24 hour lifetime. The estimated growth rate for this event was $\sim$ 3000 Mm$^2$ hr$^{-1}$, with a decay rate of $\sim$ -1000 Mm$^2$ hr$^{-1}$. EphCH 2 experienced a similar growth rate, but EphCh 3 and EphCH 4 both exhibited markedly slower growth and decay, of order 10$^2$ Mm$^2$ hr$^{-1}$ (see Table \ref{ephch_properties}). These spatial and temporal scales are reminiscent of the properties of supergranules, which typically manifest with horizontal scales of $\sim$ 30 Mm, and evolve on timescales of a few days \citep[e.g.][]{2014SoPh..289...11W, 2018LRSP...15....6R}. By contrast, traditional equatorial coronal holes can persist for weeks or months \citep[e.g.][]{2018AJ....155..153K}.

Each ephemeral coronal hole also displayed a mean magnetic flux density of a consistent polarity throughout the event lifetime. The mean magnetic flux densities were weak, only 1-2 G for EphCH 1 and EphCH 2, and 4-6 G for EphCH 3 and EphCH 4, similar to measurements of larger coronal holes \citep[e.g.][see also Section \ref{large_holes_comparison}]{2017ApJ...835..268H}. An examination of the per-pixel magnetic field also shows a `salt-and-pepper' structure, with weak magnetic field of both polarities intertwined in the coronal hole. Similar structure was observed in the measurements of large equatorial coronal holes in Section \ref{large_holes}, though both types of coronal hole are more unipolar than the surrounding quiet Sun.

No clear signatures corresponding to these events could be observed in the 1 AU solar wind data provided by the WIND spacecraft via the OMNI database. This supports the idea that these coronal holes are too compact and short-lived to have a strong impact on the solar wind that can be easily disentangled from other effects. Alternatively, in the case that the coronal hole location is not magnetically connected to the Earth, any solar wind signature would not be seen by near-Earth instrumentation. Future observations from Parker Solar Probe and Solar Orbiter at $<$1 AU may provide an promising avenue to observe the connection between such small-scale structures and the heliosphere.

In the wider context, it is uncertain whether ephemeral coronal holes are rare or simply under-reported events. A preliminary search of the SPoCA database \citep{2014A&A...561A..29V} suggests that there may be a substantial number of events with comparable areas and lifetimes, but many of these events are located near the solar limb and are the result of artificial fragmentation of larger structures. A full analysis and assessment of the systematic uncertainties in SPoCA detections is underway, but is beyond the scope of this paper. The small sample size of this study allows us to make preliminary estimates of ephemeral coronal hole characteristics. Further observations of these phenomena, particularly with SDO, will allow us to better constrain their properties, probe their physical origin, understand how they are distinct from larger polar and equatorial coronal holes, and understand their potential impact on the solar wind. 

\begin{acknowledgements}
The imaging data used in this work was provided courtesy of the NASA/SDO and the AIA and HMI science teams. The OMNI data were obtained from the GSFC/SPDF OMNIWeb interface at \url{https://omniweb.gsfc.nasa.gov}. The authors thank the Helioviewer project for providing easy browsing access to many solar imaging data sets. The authors are also grateful to the anonymous referee for numerous helpful comments that improved the final paper.
\end{acknowledgements}

\bibliographystyle{apj}
\bibliography{refs.bib}

\begin{thebibliography}{30}
\expandafter\ifx\csname natexlab\endcsname\relax\def\natexlab#1{#1}\fi

\bibitem[{{Altschuler} {et~al.}(1972){Altschuler}, {Trotter}, \&
  {Orrall}}]{1972SoPh...26..354A}
{Altschuler}, M.~D., {Trotter}, D.~E., \& {Orrall}, F.~Q. 1972, \solphys, 26,
  354

\bibitem[{{Banerjee} {et~al.}(2011){Banerjee}, {Gupta}, \&
  {Teriaca}}]{2011SSRv..158..267B}
{Banerjee}, D., {Gupta}, G.~R., \& {Teriaca}, L. 2011, \ssr, 158, 267

\bibitem[{{Boerner} {et~al.}(2014){Boerner}, {Testa}, {Warren}, {Weber}, \&
  {Schrijver}}]{2014SoPh..289.2377B}
{Boerner}, P.~F., {Testa}, P., {Warren}, H., {Weber}, M.~A., \& {Schrijver},
  C.~J. 2014, \solphys, 289, 2377

\bibitem[{{Burlaga} {et~al.}(2001){Burlaga}, {Harvey}, \& {Sheeley
  Jr}}]{2001JGR...10624915B}
{Burlaga}, L.~F., {Harvey}, K.~L., \& {Sheeley Jr}, N.~R. 2001, \jgr, 106,
  24915

\bibitem[{{Cranmer}(2009)}]{2009LRSP....6....3C}
{Cranmer}, S.~R. 2009, Living Reviews in Solar Physics, 6, 3

\bibitem[{{Feldman} \& {Widing}(2003)}]{2003SSRv..107..665F}
{Feldman}, U., \& {Widing}, K.~G. 2003, \ssr, 107, 665

\bibitem[{{Gonzalez} \& {Woods}(2002)}]{2002dip..book.....G}
{Gonzalez}, R.~C., \& {Woods}, R.~E. 2002, {Digital image processing}

\bibitem[{{Hamada} {et~al.}(2018){Hamada}, {Asikainen}, {Virtanen}, \&
  {Mursula}}]{2018SoPh..293...71H}
{Hamada}, A., {Asikainen}, T., {Virtanen}, I., \& {Mursula}, K. 2018, \solphys,
  293, 71

\bibitem[{{Hassler} {et~al.}(1999){Hassler}, {Dammasch}, {Lemaire}, {Brekke},
  {Curdt}, {Mason}, {Vial}, \& {Wilhelm}}]{1999Sci...283..810H}
{Hassler}, D.~M., {Dammasch}, I.~E., {Lemaire}, P., {et~al.} 1999, Science,
  283, 810

\bibitem[{{Heinemann} {et~al.}(2018){Heinemann}, {Temmer}, {Hofmeister},
  {Veronig}, \& {Vennerstr{\o}m}}]{2018ApJ...861..151H}
{Heinemann}, S.~G., {Temmer}, M., {Hofmeister}, S.~J., {Veronig}, A.~M., \&
  {Vennerstr{\o}m}, S. 2018, \apj, 861, 151

\bibitem[{{Hess Webber} {et~al.}(2014){Hess Webber}, {Karna}, {Pesnell}, \&
  {Kirk}}]{2014SoPh..289.4047H}
{Hess Webber}, S.~A., {Karna}, N., {Pesnell}, W.~D., \& {Kirk}, M.~S. 2014,
  \solphys, 289, 4047

\bibitem[{{Hofmeister} {et~al.}(2017){Hofmeister}, {Veronig}, {Reiss},
  {Temmer}, {Vennerstrom}, {Vr{\v s}nak}, \& {Heber}}]{2017ApJ...835..268H}
{Hofmeister}, S.~J., {Veronig}, A., {Reiss}, M.~A., {et~al.} 2017, \apj, 835,
  268

\bibitem[{{Karna} {et~al.}(2014){Karna}, {Hess Webber}, \&
  {Pesnell}}]{2014SoPh..289.3381K}
{Karna}, N., {Hess Webber}, S.~A., \& {Pesnell}, W.~D. 2014, \solphys, 289,
  3381

\bibitem[{{Kirk} {et~al.}(2009){Kirk}, {Pesnell}, {Young}, \& {Hess
  Webber}}]{2009SoPh..257...99K}
{Kirk}, M.~S., {Pesnell}, W.~D., {Young}, C.~A., \& {Hess Webber}, S.~A. 2009,
  \solphys, 257, 99

\bibitem[{{Krista} \& {Gallagher}(2009)}]{2009SoPh..256...87K}
{Krista}, L.~D., \& {Gallagher}, P.~T. 2009, \solphys, 256, 87

\bibitem[{{Krista} {et~al.}(2018){Krista}, {McIntosh}, \&
  {Leamon}}]{2018AJ....155..153K}
{Krista}, L.~D., {McIntosh}, S.~W., \& {Leamon}, R.~J. 2018, \aj, 155, 153

\bibitem[{{Krista} \& {Reinard}(2017)}]{2017ApJ...839...50K}
{Krista}, L.~D., \& {Reinard}, A.~A. 2017, \apj, 839, 50

\bibitem[{{Lemen} {et~al.}(2012){Lemen}, {Title}, {Akin}, {Boerner}, {Chou},
  {Drake}, {Duncan}, {Edwards}, {Friedlaender}, {Heyman}, {Hurlburt}, {Katz},
  {Kushner}, {Levay}, {Lindgren}, {Mathur}, {McFeaters}, {Mitchell}, {Rehse},
  {Schrijver}, {Springer}, {Stern}, {Tarbell}, {Wuelser}, {Wolfson}, {Yanari},
  {Bookbinder}, {Cheimets}, {Caldwell}, {Deluca}, {Gates}, {Golub}, {Park},
  {Podgorski}, {Bush}, {Scherrer}, {Gummin}, {Smith}, {Auker}, {Jerram},
  {Pool}, {Soufli}, {Windt}, {Beardsley}, {Clapp}, {Lang}, \&
  {Waltham}}]{2012SoPh..275...17L}
{Lemen}, J.~R., {Title}, A.~M., {Akin}, D.~J., {et~al.} 2012, \solphys, 275, 17

\bibitem[{{Murphy} {et~al.}(2018){Murphy}, {Inglis}, {Sibeck}, {Rae}, {Watt},
  {Silveira}, {Plaschke}, {Claudepierre}, \& {Nakamura}}]{2018JGRA..123.6457M}
{Murphy}, K.~R., {Inglis}, A.~R., {Sibeck}, D.~G., {et~al.} 2018, Journal of
  Geophysical Research (Space Physics), 123, 6457

\bibitem[{{Pesnell} {et~al.}(2012){Pesnell}, {Thompson}, \&
  {Chamberlin}}]{2012SoPh..275....3P}
{Pesnell}, W.~D., {Thompson}, B.~J., \& {Chamberlin}, P.~C. 2012, \solphys,
  275, 3

\bibitem[{{Reinard} \& {Biesecker}(2008)}]{2008ApJ...674..576R}
{Reinard}, A.~A., \& {Biesecker}, D.~A. 2008, \apj, 674, 576

\bibitem[{{Rincon} \& {Rieutord}(2018)}]{2018LRSP...15....6R}
{Rincon}, F., \& {Rieutord}, M. 2018, Living Reviews in Solar Physics, 15, 6

\bibitem[{{Rotter} {et~al.}(2015){Rotter}, {Veronig}, {Temmer}, \& {Vr{\v
  s}nak}}]{2015SoPh..290.1355R}
{Rotter}, T., {Veronig}, A.~M., {Temmer}, M., \& {Vr{\v s}nak}, B. 2015,
  \solphys, 290, 1355

\bibitem[{{Scherrer} {et~al.}(2012){Scherrer}, {Schou}, {Bush}, {Kosovichev},
  {Bogart}, {Hoeksema}, {Liu}, {Duvall}, {Zhao}, {Title}, {Schrijver},
  {Tarbell}, \& {Tomczyk}}]{2012SoPh..275..207S}
{Scherrer}, P.~H., {Schou}, J., {Bush}, R.~I., {et~al.} 2012, \solphys, 275,
  207

\bibitem[{{Temmer} {et~al.}(2007){Temmer}, {Vr{\v s}nak}, \&
  {Veronig}}]{2007SoPh..241..371T}
{Temmer}, M., {Vr{\v s}nak}, B., \& {Veronig}, A.~M. 2007, \solphys, 241, 371

\bibitem[{{Tsurutani} {et~al.}(1995){Tsurutani}, {Gonzalez}, {Gonzalez},
  {Tang}, {Arballo}, \& {Okada}}]{1995JGR...10021717T}
{Tsurutani}, B.~T., {Gonzalez}, W.~D., {Gonzalez}, A.~L.~C., {et~al.} 1995,
  \jgr, 100, 21717

\bibitem[{{Verbanac} {et~al.}(2011){Verbanac}, {Vr{\v s}nak}, {{\v
  Z}ivkovi{\'c}}, {Hojsak}, {Veronig}, \& {Temmer}}]{2011A&A...533A..49V}
{Verbanac}, G., {Vr{\v s}nak}, B., {{\v Z}ivkovi{\'c}}, S., {et~al.} 2011,
  \aap, 533, A49

\bibitem[{{Verbeeck} {et~al.}(2014){Verbeeck}, {Delouille}, {Mampaey}, \& {De
  Visscher}}]{2014A&A...561A..29V}
{Verbeeck}, C., {Delouille}, V., {Mampaey}, B., \& {De Visscher}, R. 2014,
  \aap, 561, A29

\bibitem[{{Wang}(2009)}]{2009SSRv..144..383W}
{Wang}, Y.-M. 2009, \ssr, 144, 383

\bibitem[{{Williams} {et~al.}(2014){Williams}, {Pesnell}, {Beck}, \&
  {Lee}}]{2014SoPh..289...11W}
{Williams}, P.~E., {Pesnell}, W.~D., {Beck}, J.~G., \& {Lee}, S. 2014,
  \solphys, 289, 11

\end{thebibliography}

\end{document}